\documentclass[manuscript,nonacm]{acmart}
\usepackage{enumitem}
\usepackage{fontawesome5}
\usepackage{tikz}
\usetikzlibrary{chains,fit,calc,positioning,calendar,arrows,shapes.geometric,quotes,babel,fit,matrix,shapes.callouts,backgrounds,arrows.meta}
\usepackage{amsmath}
\usepackage{float}
\usepackage{booktabs}
\usepackage{bm}
\usepackage{algorithmic}
\usepackage{algorithm}
\usepackage{caption}
\usepackage{wrapfig}
\usepackage{subcaption}
\usepackage{xurl}
\hypersetup{hidelinks, breaklinks=true}

\newcommand{\para}[1]{\smallskip\noindent\textbf{{#1. }}}

\newcommand{\headlesspara}[0]{\smallskip\noindent}

\begin{document}

\title[Natural Language Access Control (NLAC): From Help Desk Requests to Structured Policies]{Natural Language Access Control (NLAC):\\From Help Desk Requests to Structured Policies}

\author{Jonas Wessner}
\affiliation{%
  \institution{Ulm University}
  \city{Ulm}
  \country{Germany}}
\email{jonas.wessner@uni-ulm.de}

\author{Tobias Meuser}
\affiliation{%
  \institution{TU Darmstadt}
  \city{Darmstadt}
  \country{Germany}}
\email{tobias.meuser@kom.tu-darmstadt.de}

\author{Janek Schoffit}
\affiliation{%
  \institution{Ulm University}
  \city{Ulm}
  \country{Germany}}
\email{janek.schoffit@uni-ulm.de}

\author{Dennis Eisermann}
\affiliation{%
  \institution{Ulm University}
  \city{Ulm}
  \country{Germany}}
\email{dennis.eisermann@uni-ulm.de}

\author{Johannes Deger}
\affiliation{%
  \institution{Ulm University}
  \city{Ulm}
  \country{Germany}}
\email{johannes.deger@uni-ulm.de}

\author{Bj{\"o}rn Scheuermann}
\affiliation{%
  \institution{TU Darmstadt}
  \city{Darmstadt}
  \country{Germany}}
\email{scheuermann@kom.tu-darmstadt.de}

\author{Frank Kargl}
\affiliation{%
  \institution{Ulm University}
  \city{Ulm}
  \country{Germany}}
\email{frank.kargl@uni-ulm.de}

\renewcommand{\shortauthors}{Wessner et al.}

\begin{abstract}
  Configuring network access control policies in large, complex networks is error-prone and requires significant expert effort.
LLMs offer a promising interface for expressing such policies in natural language, but their capability for translating user requests into access policies, and the system architectures best suited to leverage LLMs, remain underexplored.
We present an architecture for natural-language access control (NLAC) that uses LLMs to translate user requests into access policies, and introduce NLACBench, a benchmark for evaluating LLM-based intent translation systems in large-scale networks.
Our evaluation across multiple state-of-the-art models shows that top-performing LLMs achieve up to 96.9\% accuracy in small-network settings, but performance degrades substantially (below 20\% for some models) as network size increases.
To address this limitation, we identify relevant network components via embedding similarity and construct compact subgraphs that are passed to the LLM.
This approach enables scaling to larger networks with up to 98.7\% accuracy, while simultaneously reducing inference time, hardware requirements, and operating costs to a constant resource budget.
Finally, a case study indicates that top-performing models exhibit largely complementary error patterns, suggesting that intent translation accuracy may be further improved through multi-LLM architectures.
\end{abstract}


\maketitle

\section{Introduction}

Network access control plays a crucial role in ensuring the security of data and IT services within organizations.
Under traditional workflows, when access control policies need to be changed (e.g., due to errors, newly joining employees, or changes in offered IT services), users typically report their issues to an IT help desk.
Network operations teams then interpret these requests, seek solutions, and deploy the necessary configuration changes.
However, as the network and user base grow, these manual processes become a bottleneck.

Intent-Based Networking (IBN)~\cite{9925251} is a recent paradigm that aims to enhance the productivity of network operations teams by introducing high-level interfaces through so-called intents.
Most IBN approaches rely on network operators to translate natural language requests from end-users into structured intents that can be processed by the IBN system.
Recent research has explored automating intent translation using techniques from Natural Language Processing (NLP) and, more recently, Large Language Models (LLMs).
However, these works assume requests are unambiguous and self-contained, which is unrealistic for inquiries written by end users.
For instance, a user may articulate their issue as ``I need access to our GitLab''.
Knowledge of a user's identity and roles is necessary to determine which GitLab instance the user is referencing.
In this paper, we explore whether LLMs can be used to correctly and reliably translate such vague requests into access policy updates.
Specifically, we investigate three research questions:

\begin{enumerate}[start=1,label={\bfseries RQ\arabic*:}, ref={RQ\arabic*}, leftmargin=*, itemsep=0.1pt]
    \item \label{rq:design} How could an LLM-powered access control architecture be designed?
    \item \label{rq:intent_translation} How well can current LLMs translate vague user requests into access policies?
    \item \label{rq:limits} What are the limitations of LLMs for intent translation and how can they be addressed? 
\end{enumerate}
To this end, we propose Natural Language Access Control (NLAC), an access control architecture that allows end users to submit policy change requests in natural language, which are then resolved to a structured format used to update the policy configuration.
To reduce the blast radius (scope of unintended side effects) of possible incorrect LLM outputs, we separate intent translation from rule-based configuration updating.
A distributed policy ownership model enables routing of change authorization to responsible service administrators, enabling efficient human oversight without centralized expert involvement.
To enable the evaluation of current and future LLMs for intent translation, we create NLACBench, a synthetic but realistic benchmark dataset spanning networks of varying scales and complexities.
The dataset comprises more than 1,000 users, 1,000 devices, 500 services, and 450 user requests paired with their intended access control policies.
We conduct an extensive study of current LLMs and find that the best-performing models achieve up to 96.9\% accuracy on NLACBench, while some open-source models show severe limitations in scaling to larger networks ($<$20\% accuracy).
To address this, we propose a network subgraph construction technique that filters LLM inputs based on multi-faceted embedding similarity.
Using this approach, we achieve scalability to larger networks while attaining up to 98.7\% accuracy at constant and substantially reduced inference costs and hardware requirements.
Finally, we study residual LLM failure modes and find that top-performing models exhibit partially complementary failure patterns, which may allow further improvement through multi-LLM cross-checking strategies.
Overall, our work makes the following contributions:
\begin{enumerate}[noitemsep]
\item An intent-based access control architecture and policy model that enable LLM-based intent translation and efficient human oversight of configuration changes,
\item a benchmark dataset to evaluate current and future LLM-based intent translation approaches,
\item an extensive study of current LLMs' potential and limitations for intent translation, and
\item a system design and implementation for scaling LLM-based intent translation to larger networks.
\end{enumerate}

\section{Related Work}

\para{Access Control Models}
The access control architecture proposed in this work is related to established models, in particular Role-Based Access Control (RBAC)~\cite{SANDHU1998237} and Attribute-Based Access Control (ABAC)~\cite{2015abac}.
RBAC grants access by assigning subjects to roles and roles to permissions, providing auditability and structural simplicity.
However, as organizations grow, RBAC suffers from role explosion: an exponential proliferation of roles required to cover every unique permission combination~\cite{roleexplosion}.
ABAC overcomes this by evaluating conjunctions of attribute predicates over subjects, resources, actions, and environmental conditions, offering substantially greater expressiveness.
However, it suffers from a complex and error-prone management process for the potentially massive and intricate rule set, which is the primary obstacle to its widespread adoption~\cite{10.1145/3007204}.
Our architecture addresses these limitations through two design decisions that enable LLM-based intent translation.
First, we separate the formulation of individual intents from system configuration through an intent-based updating algorithm, bounding both the complexity of the intent translation task and the blast radius of any mistranslation.
Second, we organize policies into administrative groups with distributed ownership, delegating authorization of policy changes to the parties most qualified to exercise it, a property that neither RBAC nor ABAC provide natively.
Our policy model is thus not a replacement for ABAC's expressiveness, but an ownership-aware, LLM-safe architectural layer that deliberately constrains the solution space to make automated, human-supervised policy management tractable at scale.

\para{Datasets for Intent Translation}
Translating natural language into structured network policies is difficult and requires rigorous evaluation to assess the effectiveness and efficiency of different approaches.
Due to the lack of publicly available real-world datasets, previous works have created their own datasets to evaluate their intent translation approaches.
This has led to a multitude of non-standardized datasets, often containing only a few samples (2--10 user requests in some works) or small illustrative scenarios (typically 1--10 network nodes) rather than use cases with realistic complexity~\cite{10.1145/3656296,10588879,10815823,sonune2025lmntoolgeneratingmachine,10835708,11029381,icissp25,cheng2025saymeannaturallanguage,modelsward25,10.1145/3764944.3764949}, leading to limited significance and poor comparability of results.
An exception is the more commonly used Lumi dataset~\cite{lumidataset}.
The Lumi dataset assumes that humans use precise natural language instructions and unique identifiers (like ``ml.cs.gitlab.uni.edu'').
Requests that require reasoning or that are sensitive to organization-specific context, such as ``our GitLab,'' are not considered.

Overall, current datasets are often insufficiently designed and standardized, and they fail to include the kind of vague problem statements that end-users would write.
To the best of our knowledge, we are the first to release a carefully designed benchmark dataset to measure the performance of LLM-based systems in solving such vague and reasoning-intensive requests.
Unlike previous works, we do not assume that requests are self-contained, but instead acknowledge that end-user requests can only be solved by taking into account dynamic information about the organization (e.g., the identity of the requesting user and the existing gitlab instances).
Consequently, our dataset is the first to include a knowledge base with information that network operators would realistically possess and can be scaled to simulate organizations of varying sizes for large-scale evaluations.

\para{Natural Language Access Control Solutions}
Several prior works have proposed approaches to translating natural language into network access policies.
However, similar to the existing intent-translation datasets, solutions such as Lumi~\cite{jacobs2021hey} and many others cannot handle vague requests such as those that would typically be written by end-users, because they assume that requests are self-contained and well-defined~\cite{10.1145/3656296,10588879,10815823,sonune2025lmntoolgeneratingmachine,10835708,11029381,icissp25,cheng2025saymeannaturallanguage,modelsward25,306009,10.1145/3672202.3673721,10855447}.
Furthermore, existing solutions are often designed to operate only in small-scale demonstrations and do not scale to large networks.
Specifically, many prior works that use LLMs prompt the model to output the entire global network configuration directly, which is evaluated only in simple test cases where small networks are configured from scratch, but is unrealistic for managing large networks with complex configuration states~\cite{10.1145/3656296,10.1145/3764944.3764949,10588879}.
Existing research further lacks a comprehensive evaluation of recent LLMs for understanding and solving reasoning-intensive network access control tasks, as previously published works evaluate only a few, and often significantly outdated, LLMs (e.g., GPT-2, GPT-3.5, BERT base)~\cite{10310190,10.1145/3764944.3764949,sonune2025lmntoolgeneratingmachine,10.1145/3656296,11029381,icissp25,10.1145/3764944.3764949}.
Building on the NLAC architecture and the NLACBench dataset, our work addresses these shortcomings by designing and benchmarking an intent translation system that solves vague requests in realistic and large-scale environments.

\section{Foundations}

In this section, we define the necessary terminology and identify characteristics of access control management that guide our design decisions.

\subsection{Terminology}\label{sec:foundations:terminology}

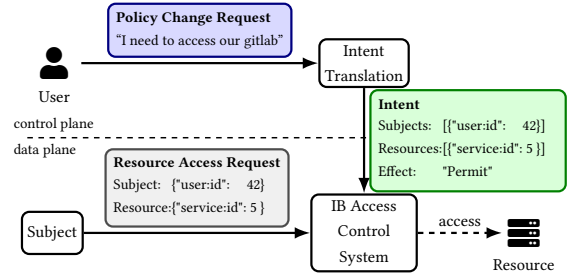
\begin{wrapfigure}[11]{b}{0.5\linewidth}
	\centering
	\resizebox{\linewidth}{!}{%
		\usetikzlibrary{positioning, shapes.geometric, arrows.meta, fit}
	\begin{tikzpicture}[
		roundbox/.style={very thick, rectangle, draw, rounded corners, minimum width=1cm, minimum height=0.8cm, align=center},
		box/.style={rectangle, draw, minimum width=1cm, minimum height=0.8cm, align=center},
		smallbox/.style={font=\small,rectangle, draw, rounded corners, minimum width=3cm, align=center, inner sep=0.15cm},
		bigbox/.style={rectangle, draw, rounded corners, minimum width=3cm, minimum height=2.5cm, align=center},
		db/.style={cylinder, draw, shape aspect=0.5, aspect=0.2, cylinder uses custom fill, cylinder body fill=white,
				cylinder end fill=white, align=center, minimum width=0.85cm},
		dbvert/.style={cylinder, draw, shape border rotate=90, aspect=0.5, aspect=0.2, cylinder uses custom fill, cylinder body fill=white,
				cylinder end fill=white, align=center},
		arrow/.style={-{Latex}, very thick},
		arrowdoubleheaded/.style={{Latex}-{Latex}, very thick},
		idx/.style={circle,draw=black,fill=white,line width=1px,inner sep=1pt,minimum size=1mm,minimum width=1mm,minimum height=1mm},
		node distance=1.2cm,
		]

		\node[roundbox, minimum width=2.0cm] (acs) {IB Access\\Control\\System};

		\node[roundbox, left=4.3cm of acs] (subject) {Subject};
		\node[text width=1.2cm, right=1.3cm of acs, align=center,label=below:Resource] (resource) {\Huge\faServer};

		\node[roundbox, above=2.0cmof acs,  align=center] (intenttranslation) {Intent\\Translation};
		\node[text width=0.8cm, left=4.5cm of intenttranslation, align=center,label=below:User] (user) {\Huge\faUser};

		\draw [dashed, thick] ($(user.south west)+(0,-1cm)$) -- ($(user.south west)+(10cm,-1cm)$) node[pos=0.05, align=left] () {\small control plane\\\small data plane};

		\draw[arrow] (subject) -- (acs) node[midway, above, smallbox, yshift=0.1cm, align=left, fill=gray!12, draw=black!70] (rarbox)
		{
			\textbf{Resource Access Request}\\
			\begin{tabular}{@{}l@{}l@{}c@{}l@{}}
				Subject:  & \{"user:id":    & 42 & \} \\
				Resource: & \{"service:id": & 5  & \}
			\end{tabular} };

		\draw[arrow, dashed] (acs) -- ($(resource.west)+(0.3,0cm)$) node[midway,above]{access};

		\draw[arrow] (intenttranslation.south) -- (acs.north) node[midway, right, smallbox, xshift=0.1cm,  fill=green!15, draw=green!50!black,align=left] (intentbox)
		{\textbf{Intent \phantom{Increases h. Spacing}} \\
			\begin{tabular}{@{}l@{}l@{}c@{}l@{}}
				Subjects:  & [\{"user:id":    & 42 & \}] \\
				Resources: & [\{"service:id": & 5  & \}] \\
				Effect:    & "Permit"         &    &
			\end{tabular}};

		\draw[arrow] (user) -- (intenttranslation) node[midway, above, smallbox, yshift=0.1cm, fill=blue!15, draw=blue!50!black, align=left] (pcrbox)
		{\textbf{Policy Change Request}\\ ``I need to access our gitlab'' };

	\end{tikzpicture}
	}
	\caption{Intent-based access control system overview.}
	\label{fig:terminology}
\end{wrapfigure}

Figure \ref{fig:terminology} shows a high-level overview of an intent-based access control system.

\emph{Subjects}, such as user accounts or application processes, access information or functionality provided by \emph{resources} (e.g., applications or hardware resources) (cf. NIST \cite{stafford2020zero}).

\emph{Resource Access Requests~(RARs)} are attempts by subjects to access resources~\cite{10310190,stafford2020zero}.
For instance, if a user queries a remote database, this results in an RAR that is checked by the intent-based access control system, which either allows or denies the access.
RARs are characterized by several attributes, including subject ID, resource ID, and other optional attributes (e.g., device execution context).
Figure~\ref{fig:terminology} illustrates an RAR as a \textit{request} in simplified eXtensible Access Control Markup Language~(XACML)~\cite{OASIS_XACML_3_0} notation.

\emph{Policy Change Requests~(PCRs)}, in this paper, refer to vague problem statements or requirements expressed in natural language by an end user, such as those that would arrive at IT help desks; for example, \emph{``I need to access our gitlab''}.

\emph{Intents} are a disambiguated, machine-readable form of PCRs.
They consist of a scope (i.e., which RARs they apply to) and an intended effect (permit/deny access).
Intents are \emph{control-plane} artifacts: users express an intended change in natural language (PCR), which is translated into a machine-readable intent that is used to update the access control configuration.
RARs, by contrast, are \emph{data-plane} events that occur at runtime when a subject attempts to access a resource, and are permitted or denied based on the current configuration.

\subsection{Design Principles}\label{sec:foundations:design_principles}

To derive essential requirements for an intent-based access control system, we conducted interviews with network security experts and network operations teams from three universities.
The interviews focused on recurring challenges in access control policy management, and were complemented by an analysis of a dataset comprising 100 real-world access control-related help desk inquiries.
From these insights, we distilled the following three challenges to guide our design decisions:

\para{Vague and Diverse Language}
Since PCRs can be submitted by users from different professional backgrounds, the style and level of detail may vary significantly.
While technical users can often precisely explain their requirements, most user requests are vague and ambiguous when taken out of context.
For example, ``our gitlab'' is ambiguous if the identity of the user is not considered.  

\para{Conflicts and Intent Evolution}
As intents accumulate in the system over time, conflicts between intents will arise and must be resolved to accommodate evolving operational needs.
For instance, department A might want to deny external access to their gitlab, but later an external collaborator may request access. This conflict must be resolved, e.g., through an exception to the first rule.

\para{Blast Radius}
Access control configuration changes may cause unintended side effects that can result 
in system downtimes or security vulnerabilities.
In established models such as RBAC and ABAC, there is no architectural mechanism that 
scopes the effect of a change to a particular administrative domain, such that a single 
modified rule can silently affect an unbounded number of subject-resource pairs across 
the entire system~\cite{10.1145/3007204}.

\section{Natural Language Access Control (NLAC)}
\label{sec:architecture}

\begin{figure*}[htb]
    \centering
    \resizebox{0.9\linewidth}{!}{%
    \begin{tikzpicture}[
	box/.style={very thick, rectangle, draw, rounded corners, minimum width=3cm, minimum height=0.8cm, align=center},
	smallbox/.style={rectangle, draw, rounded corners, inner sep= 0.15cm, align=center},
	bigbox/.style={very thick, rectangle, draw, rounded corners, minimum width=3cm, minimum height=2.5cm, align=center},
	db/.style={cylinder, draw, shape aspect=0.5, aspect=0.2, cylinder uses custom fill, cylinder body fill=white,
			cylinder end fill=white, align=center, minimum width=0.85cm},
	dbvert/.style={cylinder, draw, shape border rotate=90, aspect=0.5, aspect=0.2, cylinder uses custom fill, cylinder body fill=white,
			cylinder end fill=white, align=center},
	arrow/.style={-{Latex}, very thick},
	arrowdoubleheaded/.style={{Latex}-{Latex}, very thick},
	idx/.style={circle,draw=black,fill=white,line width=1px,inner sep=1pt,minimum size=1mm,minimum width=1mm,minimum height=1mm},
	node distance=1.2cm,
	highlighted/.style={fill=pink!70},
	smalltext/.style={font=\small,},
	]

	\node[box, minimum width=4.1cm] (pep) {Policy Enforcement Point};

	\node[text width=1cm,left=3.5cm of pep, align=center,label=below:System] (system) {\Huge\faLaptop};
	\node[box, left=of system, minimum width=2cm] (subject) {Subject};
	\node[text width=1.2cm, right=2.35 of pep, align=center,label=below:Resource] (resource) {\Huge\faServer};

	\node[bigbox, dashed, above=1.4cm of pep, minimum width=5.3cm, minimum height=3.7cm] (pdp) {};

	\node[box, above=0.1cm of pdp.south, xshift=1.235cm, minimum height=1.0cm, minimum width=2.35cm] (policyadministrator) {Policy\\Administrator};
	\node[box, left=0.1cm of policyadministrator, minimum height=1.0cm, minimum width=2.35cm] (pe) {Policy\\Engine};

	\node[smallbox, above=1.2cm of pdp.south, minimum height=0.85cm, thick, anchor=south, highlighted, minimum width=5.1cm] (ipc) {\textbf{Intent Policy Configuration}};
	\node[smallbox, minimum height=0.85cm, above=0.1cm of ipc.north, thick,  minimum width=5.1cm, highlighted] (policyengine) {\textbf{Intent Configuration Manager}};
	\node[text width=1.2cm, right=2.2cm of policyengine, align=center,label=below:Policy Owners] (policyowners) {\Huge\faUsers};

	\node[box, left=1.8cm of policyengine, thick, align=center, highlighted] (intenttranslation) {\textbf{Intent}\\\textbf{Translation}};
	\node[text width=0.8cm, left=2.5cm of intenttranslation, align=center,label=below:User] (user) {\Huge\faUser};
	\node[below=0.1cm of pdp.north] (pdplabel) {Policy Decision Point};

	\node[text width=1.5cm, below=of intenttranslation, align=center,label=below:Knowledge Base] (knowledgebase) {\Huge\faDatabase};

	\draw[arrow] (subject) -- (system);
	\draw[arrow] (system) -- (pep) node[midway, above, align=center, smallbox, yshift=0.1cm, fill=black!12, draw=black!70]{Resource Access\\Request (RAR)};
	\draw[arrow, dashed] (pep) -- (resource) node[midway,above, smalltext]{access};

	\draw[arrowdoubleheaded] (policyengine.east) -- (policyowners.west) node[smalltext, midway,above,align=center, xshift=0.1cm]{authorize\\changes};
	\draw[arrow] (intenttranslation.east) -- (policyengine.west) node[midway, above, smallbox, yshift=0.1cm, fill=green!15, draw=green!50!black]{Intent};
	\draw[arrow] (user) -- (intenttranslation) node[smalltext, midway, above, align=center, smallbox, yshift=0.1cm, fill=blue!15, draw=blue!50!black]{Policy Change\\Request (PCR)};
	\draw[arrow] (knowledgebase) -- (intenttranslation) node[midway, right, align=left, smalltext]{Context\\Information};

	\draw[arrowdoubleheaded] (pep) -- (pdp) node[midway, below, sloped, smalltext]{decide};

\end{tikzpicture}
    }
    \caption{NLAC architecture based on the architecture by NIST~\cite{stafford2020zero} with contributions highlighted in red.}
    \label{fig:architecture}
\end{figure*}
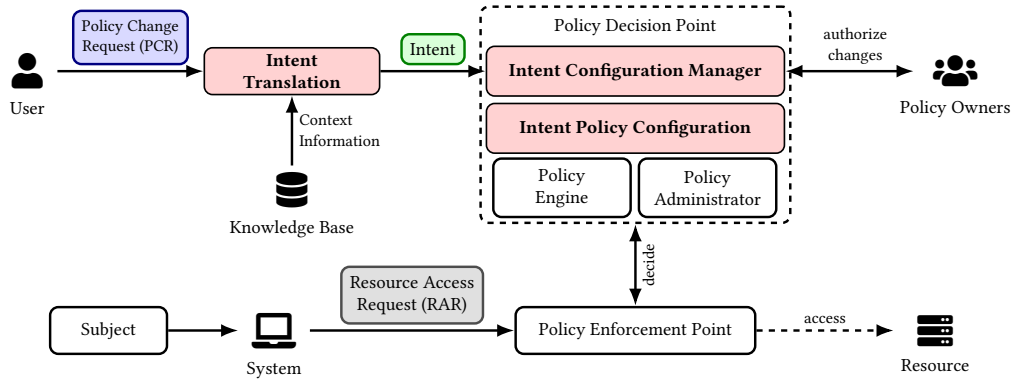

\noindent
We present Natural Language Access Control (NLAC), an intent-based access control architecture based on the zero trust architecture proposed by the National Institute of Standards and Technology (NIST)~\cite{stafford2020zero}, as illustrated in Figure~\ref{fig:architecture}.
NLAC extends the original NIST architecture to enable end users to configure access policies directly through natural language.
To this end, we propose an \emph{intent translation} module that functions as an intermediary between non-technical users and the system, similar to a help desk in classical workflows.
The intent translation module considers contextual information retrieved from a knowledge base to disambiguate Policy Change Requests (PCRs) and translate them into structured intents.
Translated intents are stored in the so-called \emph{Intent Policy Configuration (IPC)}, a policy model that organizes policies in administrative groups that are linked to policy ownership.
The \emph{intent configuration manager} defines the logic for updating the IPC for a given translated intent.
To maintain control over policy changes, the owners of the affected policy groups are asked for authorization, resulting in a distributed and scalable authorization mechanism.

In the remainder of this section, we present our architecture in more detail.
Throughout the section, we use the following \textbf{running example}: user~101 from the IoT lab submits the PCR \textit{``I need to do experiments with the robot arm''}, which should be translated into an intent granting user~101 access to the robot arm controller service (ID~107) in the robotics lab.
Additionally, we present a full step-by-step walkthrough of this example in Appendix~\ref{app:processing_of_pcrs}.

\subsection{Formal Definitions for Resource Access Requests and Intents}

We define a Resource Access Request (RAR) $r$ as follows:

\[
    r = (a_1 \dots, a_n)
    \in A_1 \times \cdots \times A_n,
\]
where each $A_i$ is the attribute space containing possible values for the $i$-th attribute of the RAR, and $n$ is a fixed system parameter.
RARs have at least two attributes, namely the subject sending this RAR (e.g., user ID~101) and the target resource (e.g., service ID~107).

We define an intent $(\sigma, \delta)$ as a tuple consisting of a selector $\sigma$ and a decision $\delta$.
The selector $\sigma$ determines which RARs this intent should affect (e.g. all RARs that come from the user with ID~101 and access the resource with ID~107), and the decision $\delta$ determines the expected system behavior for these RARs (allow or deny).
A selector is a tuple of $n$ predicates $S_i$ defined on the corresponding attributes of RARs:
\begin{align*}
    \sigma & = (S_{1}, \dots, S_n),
    \quad  S_i     : A_i \to \{\text{true},\text{false}\}
    \text{ for all } i \in \{1, \dots, n\}.
\end{align*}
The predicates express conditions for attribute values to match an RAR. For instance:

\begin{itemize}[noitemsep]
    \item if $A_i$ is a finite set (e.g., the set of subject IDs), we can define a selector that matches a subset. For example, $S_i(a_i) = \text{true}\; \text{if}\; a_i \in \{101\}\; \text{else}\; \text{false}$ to only match RARs from subject with ID~101;
    \item if $A_i$ is an ordered infinite set (e.g., timestamps), range selectors can be defined: $S_i(a_i) = \text{true} \; \text{if} \; t^{\min} \le a_i \le t^{\max} \; \text{else} \; \text{false}$;
    \item and wildcards can be defined as: $S_i(a_i) = \text{true}$
\end{itemize}
A selector matches an RAR ($\sigma \triangleright r$) if all of its predicates match the corresponding RAR attributes:
\[
    \sigma \triangleright r
    \;\;\Leftrightarrow\;\;
    \forall i \in \{1, \dots, n\},\; S_i(a_i) = \text{true}.
\]
The decision $\delta \in \{0, 1\}$ indicates the intended decision for all $r$ that $\sigma$ matches, where $\delta=1$ indicates to permit and $\delta=0$ to deny access.

\subsection{Context-aware Intent Translation}
\label{sec:architecture:intent_translation}

The intent translation module provides a natural language interface for end users to express their access policy needs.
It aims to automate the process of understanding PCRs and translating them into a machine-readable format, modeled as intents $(\sigma, \delta)$.
In the NLAC architecture, intent translation is intentionally separated from authorization and configuration updating to manage the risk of incorrect LLM outputs.
Additionally, this design simplifies the task for the LLM, as it does not need to understand the current access policy configuration.

PCRs can be vague and ambiguous when taken out of context.
To resolve vagueness in PCRs (e.g., finding the service needed to control the ``robot arm''), the intent translation module must be aware of organization-specific information, such as user identities and service names.
We model this as a knowledge base that contains information about subjects and resources as well as their relationships.
A practical implementation could query existing data sources such as NetBox~\cite{netbox} and LDAP~\cite{ldap}.
Our system design of an LLM-based intent translation module is presented in Section~\ref{sec:prototype} and evaluated in Section~\ref{sec:eval}.

\subsection{Intent Policy Configuration}\label{sec:architecture:ipc}

The Intent Policy Configuration (IPC) is a control-plane policy model that introduces distributed ownership and facilitates automatic intent-based updating and efficient human supervision.
Policies are organized in administrative groups that express the intended ingress and egress policies for that group.
This way, the owners of each group maintain control over their own policies, and accesses between groups are permitted only when both groups' policies agree.
 
To formally define the IPC, we first divide the set of all subjects and resources into $m$ administrative groups $g_1$ to $g_m$ which define ownership.
For instance, each department in an organization could be treated as such a group and delegate one member as policy owner.
The IPC expresses policies as sets of selectors.
These selector sets constitute whitelists of allowed accesses: RARs matched by at least one selector in a selector set are considered allowed under that selector set.
To revoke access, existing selectors are modified or removed from the whitelists.
More specifically, the IPC is organized in $m$ ingress selector sets $\Pi_1^\mathrm{in}$ to $\Pi^\mathrm{in}_m$ and $m$ egress selector sets $\Pi^\mathrm{out}_1$ to $\Pi^\mathrm{out}_m$.
The ingress selector set $\Pi^\mathrm{in}_i$ and egress selector set $\Pi^\mathrm{out}_i$ define the allowed incoming and outgoing RARs for group $i$.
For a given RAR $r$, let $s$ and $d$ be the indices of the groups owning the subject and resource, respectively.
Then, $r$ is granted if there exists at least one selector in the egress set of group $g_s$ and at least one selector in the ingress set of group $g_d$ allowing the access:
\[
    \exists \sigma_{\mathrm{out}} \in \Pi_s^\mathrm{out},\,
    \exists \sigma_{\mathrm{in}} \in \Pi_d^\mathrm{in} : 
    (\sigma_{\mathrm{out}} \triangleright r) \wedge ( \sigma_{\mathrm{in}} \triangleright r).
\]

This formulation ensures that RARs are only granted if both administrative groups explicitly allow it, capturing the mutual consent of responsible parties.
For instance, an RAR from user~101 to access the robot arm controller service is only granted if the egress selector set of the IoT lab and the ingress selector set of the robotics lab both match it.

\subsection{Intent Configuration Manager}\label{sec:architecture:icm}

The Intent Configuration Manager defines the logic for updating the selector sets of the IPC for a given intent.
It further manages the authorization of proposed policy changes.

\para{Intent Resolution}
For a given intent $(\sigma, \delta)$, the intent configuration manager determines how the selector sets in the IPC must be modified so that the configuration aligns with the intent.
It first identifies the two affected selector sets $\Pi^\mathrm{out}_{s}$ and $\Pi^\mathrm{in}_{d}.$\footnote{For simplicity, we assume that each change request affects only a single source and destination group. However, the same approach also applies to scenarios involving multiple groups by executing the update algorithm for each affected group.}
In our example, these are the egress selector set of the IoT lab and the ingress selector set of the robotics lab.
For both selector sets, we then compute their required updates using Algorithm~\ref{algo:update_sets}.
The proposed configuration changes then proceed to the policy owners for review.

\begin{algorithm}[b]
    \renewcommand{\algorithmiccomment}[1]{\hfill{// #1}}
    \begin{algorithmic}[1]
        \STATE \textbf{Function} \textsc{UpdateSelectorSet}(($\sigma, \delta), \Pi$)
        \IF{$(\delta = 1) \land \lnot(\exists \sigma' \in \Pi \mid \sigma \subseteq \sigma')$}
        \STATE $\Pi \gets \Pi \setminus \{\sigma' \in \Pi \mid \sigma' \subseteq \sigma\}$ \hfill  \COMMENT{remove redundancies}
        \STATE $\Pi \gets \Pi \cup \{\sigma\}$ \hfill \COMMENT{add new selector\phantom{--------}}
        \ENDIF
        \IF{
            $(\delta = 0) \land (\exists r \in A_1 \times \dots \times A_n, \exists \sigma' \in \Pi
            \newline
            \hspace*{2em}%
            \mid (\sigma' \triangleright r) \land (\sigma \triangleright r))$
        }

        \FORALL{$\sigma' \in \Pi_{\mathrm{old}}$ where $\Pi_{\mathrm{old}} \gets \Pi$}
        \STATE $\Pi \gets \Pi \setminus \{\sigma'\}$ \COMMENT{remove old selector\phantom{--}}
        \STATE $\Pi \gets \Pi \cup \mathrm{exclude}(\sigma', \sigma)$ \COMMENT{add new selector\phantom{--------}}
        \ENDFOR
        \ENDIF
        \STATE \textbf{return} $\Pi$
    \end{algorithmic}
    \caption{Update selector set $\Pi$ based on intent $(\sigma, \delta)$.}
    \label{algo:update_sets}
\end{algorithm}

The updating algorithm takes an intent $(\sigma, \delta)$ and a selector set $\Pi$ as input and returns the updated selector set.
We use $\sigma \subseteq \sigma'$ to denote that $\sigma$ is covered by $\sigma'$, i.e., $\sigma \triangleright r \Rightarrow \sigma' \triangleright r$ for all RARs $r \in A_1 \times \dots \times A_n$.
Further, we use $\mathrm{exclude}(\sigma', \sigma)$ to denote a subroutine that returns a set of selectors matching all RARs that $\sigma'$ matches, except for those that $\sigma$ matches (see Appendix~\ref{app:exclude_def} for a formal definition).

If the intended decision is to permit access ($\delta = 1$), 
we check whether $\sigma$ is already covered by a selector in $\Pi$.
If not, we add $\sigma$ to the set and remove any redundant selectors (those that are covered by the new selector).
If the intended decision is to deny access ($\delta = 0$), we check if there is any overlap between $\sigma$ and any of the selectors in $\Pi$.
If so, we modify the selectors in $\Pi$ accordingly to exclude RARs in the scope of $\sigma$.

\para{Change Authorization}
While NLAC automates intent translation and resolution, human administrators retain control over authorizing proposed changes.
Instead of a central network operations team that interprets and implements PCRs, NLAC delegates authorization to the policy owners of the affected selector sets.
For instance, the policy owners of the IoT lab and of the robotics lab are asked to confirm the access from user~101 to the robot arm controller service.
We make this design decision because policy owners are more qualified to make these decisions than a central network operations team: for example, the administrator of the robotics lab knows best whether user~101 from the IoT lab should be granted access to their services.
This concept of distributed change reviewing also scales with network size, as organizations are naturally divided into departments and working groups of manageable size.

\para{Implications}
First, designing policies as unordered whitelists simplifies automated intent-based updating, enabling separation of LLM-based intent translation from the rule-based modification of the policy configuration.
Second, as one intent can cause changes to at most two selector sets, the blast radius of misconfigurations and the resulting risk of lateral movement are reduced.
Third, maintaining ownership of policies allows for efficient and scalable human oversight and reduces the need for expert involvement.
Finally, if LLMs can be reliably used for intent translation, NLAC leads to a largely automated access control management workflow.

\section{Designing an LLM-based Intent Translation System}\label{sec:prototype}

\begin{figure}[htb]
  \centering
  \resizebox{0.9\linewidth}{!}{
  \input{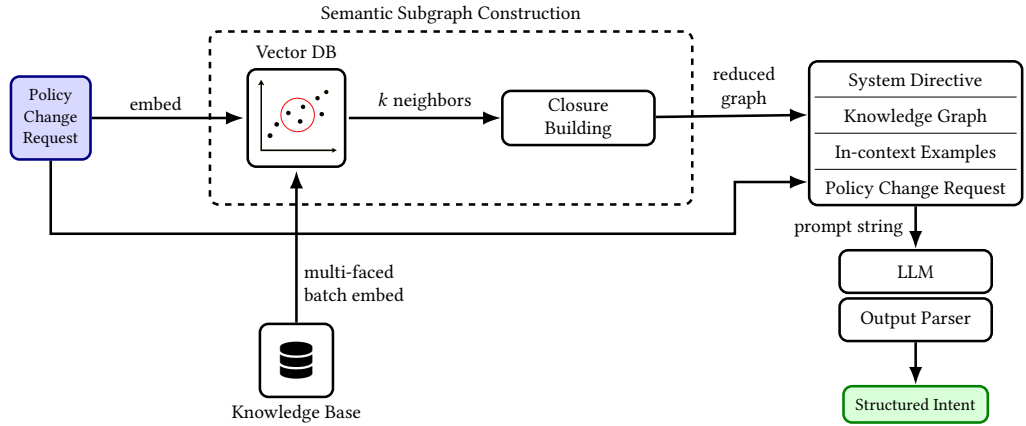}
  }
  \caption{LLM-based intent translation design.}
  \label{fig:design}
\end{figure}

In this section, we propose a design for an LLM-based intent translation system (Figure~\ref{fig:design}) within the Natural Language Access Control (NLAC) architecture, which we use in Section~\ref{sec:eval} to evaluate the intent translation abilities of state-of-the-art LLMs.
The intent translation system takes two inputs: the natural-language Policy Change Request (PCR) and a knowledge base, which contains information about the network and the organization.
Both inputs are integrated into a prompt, which is passed to the LLM.
The LLM is instructed to translate the request into a structured intent (e.g., in JSON), which is subsequently checked for syntactic correctness by an output parser.

\para{Semantic Subgraph Construction}
To assess the raw performance of current LLMs, we provide a baseline implementation where the knowledge base is passed directly to the LLM.
However, excessively large contexts are undesirable, since for standard transformer-based LLMs, computation time and VRAM usage increase quadratically with input size~\cite{NIPS2017_3f5ee243}.
This has implications for system requirements in on-premise deployments and, likewise, for remote API inference costs (usually billed per token~\cite{openai_api_pricing}).
Furthermore, our scaling experiment in Section~\ref{sec:eval:exp3} indicates that some LLMs struggle with intent translation when the provided context is too large, which aligns with general research on LLMs' ability to handle long-range context dependencies~\cite{lietal2024loogle}.
To address these limitations, we introduce a semantic network subgraph construction approach that aims to reduce the amount of context information shown to the LLM by finding relevant entities in the knowledge base for a given PCR.

Our subgraph construction approach builds on RAG~\cite{2020_RAG} to retrieve entities, such as users, devices, and services, that are semantically similar to the PCR from a vector database.
To this end, we define functions that map each entity in the knowledge base to a text string (e.g., for users, the string contains the user's name and their roles).
We then precompute embedding vectors for all entities using a text embedding model (e.g., text-embedding-3-small) and store them in a vector database for efficient retrieval.\footnote{Computing an embedding of a relatively short text adds negligible overhead compared to the subsequent call to the LLM.}
By supplying multiple mapping functions, each entity can be described using different facets of its semantics.
Each such description yields a different embedding vector that can be used as an independent search index.
For instance, a service can be mapped to a high-level description of the functionality it provides, as well as a low-level description including attributes such as its IP address and other technical information.
In this way, PCRs written at different abstraction levels can index the same entities based on different semantics.
For a given PCR, we then compute its embedding on the fly and look up the $k$ nearest neighbors of each entity type in the vector database.
The hyperparameter $k$ can be optimized independently per entity type.

After retrieval, the selected entities do not yet form a coherent subgraph.
For instance, a service may be retrieved without the device hosting it.
Preliminary experiments showed that LLMs produce incorrect results when confronted with such missing entities.
To address this, we build a 1-step closure around the entities retrieved from RAG to construct a coherent graph:
For every retrieved entity, we add all entities that it directly references to the subgraph (e.g., the device that hosts a retrieved service, or all devices that a retrieved user owns).
This resolves dangling references from entities retrieved based on semantic similarity and place them in the context of their neighbors in the graph.

In Section~\ref{sec:eval}, we analyze the possible performance gains and cost reductions when using subgraph construction in comparison to using standalone LLMs for intent translation.

\para{Prompt}
The \emph{system directive} contains a description of the task, explaining that the LLM receives a user request and is asked to construct an intent.
It also includes instructions about the expected output format (e.g., JSON representation of XACML) and an example output.
To allow the LLM to resolve references in natural language (e.g., ``file server'') to unique identifiers (e.g., resource ID 5), we integrate a JSON representation of the \emph{knowledge graph} into the prompt.
We further add \emph{in-context examples} of PCRs and their expected intents to the prompt.
Such examples can be taken from a small, manually verified collection of previously processed PCRs and their resulting intents, augmented by an explanation of how each intent was constructed.
Finally, we add the PCR received from the user and ask the LLM to construct the intent.
The exact prompt template used in our implementation can be found in Appendix~\ref{app:prompts}.

\section{NLACBench: Benchmark Dataset}\label{sec:dataset}

To explore whether LLMs can be used for intent translation in real-world scenarios, it is crucial to quantify their performance on a realistic benchmark dataset.
In this section, we design and create NLACBench, a benchmark dataset for evaluating intent translation systems, covering 450 Policy Change Requests (PCRs) in natural language paired with their expected intents and a scalable knowledge base that allows simulating PCR resolution in organizations of varying sizes.
The dataset is publicly available on GitHub~\cite{artifacts_github}, providing a foundation for comparing current and future LLMs and intent translation approaches.

We began by interviewing network operators to leverage their firsthand experience with policy change requests at IT help desks, using this input to design a schema for the knowledge base comprising relevant entity types for resolving common PCRs.
We identified four essential entity types: \texttt{user}, \texttt{role}, \texttt{device}, and \texttt{service}, along with their relationships, as shown in Figure~\ref{fig:data_layout}.

Next, we define a schema for PCRs and their corresponding intents.
PCRs consist of a natural-language text, similar to messages received at IT help desks, and the ID of the requesting user.
Intents consist of a subject ID (either a user or a service), a resource ID (a service), a list of permitted source devices (source execution contexts), an effect (permit or deny), and two optional attributes: expiry date and time of day.

Using this data model, we then created a knowledge base and, within this environment, pairs of policy change requests and intents.
Collecting real-world data (e.g., from help desk email inboxes) for a public benchmark dataset is challenging because policy change requests can contain arbitrary personal and sensitive information, such as names, email addresses, or passwords.
Therefore, we decided to generate the dataset from synthetic data.

\subsection{Generation of the Knowledge Base}

To quantify intent translation performance in the context of large organizations, the knowledge base (cf. Figure~\ref{fig:data_layout}) must be populated with a large number of entities, rendering manual data creation infeasible.
At the same time, rule-based data generation approaches fail to produce sufficiently diverse outputs, such as realistic constellations of departments and offered IT services.
Large Language Models (LLMs) therefore appear to be a promising alternative given their strong creative writing capabilities~\cite{gomez2023confederacy}.
However, using LLMs to generate datasets introduces several challenges:

\para{Limited context window}
When generating large datasets, previously generated data increasingly occupies the available context window.
Due to the fixed context size of LLMs and their limited ability to model long-range dependencies~\cite{hosseini2025efficient,lietal2024loogle}, this increases the likelihood of inconsistencies or redundancies between newly generated and previously generated data points.

\para{Interdependencies}
Entities in the knowledge base are interconnected by relationships (e.g., users can own devices).
Generating the knowledge base in a single LLM call can lead to arbitrary dependencies between entities.
Yet to quantify the scalability of intent translation systems to larger organizations, the knowledge base should be modular and configurable in size.

\begin{figure}
	\centering

	\begin{minipage}{0.49\textwidth}
		\centering
		\resizebox{!}{0.6\linewidth}{
			\begin{tikzpicture}[
        entity/.style={anchor=north west, draw, thick, align=left, inner sep=0.3cm},
        arrow/.style={-Stealth, thick},
        node distance=1.5cm and 1.5cm
    ]

    \usetikzlibrary{shapes.geometric, arrows.meta, positioning}

    \node (USER) [entity] {
        \begin{minipage}[t][1.7cm][t]{1.8cm}
            \textbf{User}\\[0.1cm]
            {
            \small
            – name\\
            – department
            }
        \end{minipage}
    };

    \node (ROLE) [entity, below=of USER] {
        \begin{minipage}[t][1.7cm][t]{1.8cm}
            \textbf{Role}\\[0.1cm]
            {
            \small
            – title\\
            – description
            }
        \end{minipage}
    };

    \node (DEVICE) [entity, right=of USER] {
        \begin{minipage}[t][1.7cm][t]{1.8cm}
            \textbf{Device}\\[0.1cm]
            {
            \small
            – type\\
            – IP addr.
            }
        \end{minipage}
    };

    \node (SERVICE) [entity, below=of DEVICE] {
        \begin{minipage}[t][1.7cm][t]{1.8cm}
            \textbf{Service}\\[0.1cm]
            {
            \small
            – name\\
            – description\\
            – protocol
            }
        \end{minipage}
    };

    \draw[arrow] (USER) --  node[left, align=right] {0..n\\has\\[0.0cm]1} (ROLE);
    \draw[arrow] (USER) -- (DEVICE) node[midway, align=center] {owns\\[0.5em]0..1\phantom{---}0..n};
    \draw[arrow] (DEVICE) -- node[left, align=right] {1\\provides\\0..n} (SERVICE)  ;

    \draw ($(USER.north west)-(0,0.7cm)$) -- ($(USER.north east)-(0,0.7cm)$);
    \draw ($(DEVICE.north west)-(0,0.7cm)$) -- ($(DEVICE.north east)-(0,0.7cm)$);
    \draw ($(ROLE.north west)-(0,0.7cm)$) -- ($(ROLE.north east)-(0,0.7cm)$);
    \draw ($(SERVICE.north west)-(0,0.7cm)$) -- ($(SERVICE.north east)-(0,0.7cm)$);

\end{tikzpicture}
		}\caption{NLACBench knowledge base schema.}\label{fig:data_layout}
	\end{minipage}
	\hfill
	\begin{minipage}{0.49\textwidth}
		\centering
		\resizebox{!}{0.6\linewidth}{
			\usetikzlibrary{arrows.meta}
\begin{tikzpicture}[
	node distance=0.5cm and 0.8cm, inner sep=0.2cm,
	box/.style={very thick, rectangle, rounded corners, draw=black, minimum width=2.7cm, align=center, minimum height=0.8cm},
	colorbox/.style={box, fill=purple!15},
	arrow/.style={-{Latex}, very thick},
	dot/.style={circle, fill, inner sep=1pt}
	]

	\node[box, rounded corners] (domain) {\textbf{domain}\\university};
	\node[colorbox, below=of domain] (llm1) {\textbf{LLM}\\decomposition};
	\node[box, below=of llm1, rounded corners] (subdomain1) {\textbf{subdomain 1}\\department X};

	\node[box, right= of subdomain1, rounded corners] (subdomain2) {\textbf{subdomain n}\\ (...)};

	\node[colorbox, below=of subdomain1] (llm2) {\textbf{LLM}\\expansion};
	\node[box, right= of llm2, rounded corners] (facts) {\textbf{facts:}\\unique names and
		IDs};
	\node[box, below=of llm2, rounded corners] (output) {\textbf{data}\\department info};

	\node[dot] at ($(subdomain1.east)!0.5!(subdomain2.west)$) (d2) {};
	\node[dot, right=0.1cm of d2] (d1) {};
	\node[dot, left=0.1cm of d2] (d3) {};

	\draw[arrow] (domain) -- (llm1);
	\draw[arrow] (llm1) -- (subdomain1);
	\draw[arrow] (llm1) -- (subdomain2);
	\draw[arrow] (subdomain1) -- (llm2);
	\draw[arrow] (facts) -- (llm2);
	\draw[arrow] (llm2) -- (output);

\end{tikzpicture}
		}\caption{Synthesizing the NLACBench knowledge base.}
		\label{fig:taxonomic_generation}
	\end{minipage}
\end{figure}

\headlesspara
To address these challenges, we propose a method for synthetic data generation that builds on the principles of dataset-wise decomposition and multi-step generation~\cite{long2024llms,wan2023explore,li2024synthetic}.
As illustrated in Figure~\ref{fig:taxonomic_generation}, we use multiple LLM calls to produce synthetic data:
First, we divide the target domain (e.g., a campus network) into smaller subdomains, such as departments, and then expand each subdomain into a set of detailed entities.
In the first generation step, the model is prompted to produce a list of diverse subdomains within the given domain, together with brief descriptions that provide the necessary context for the subsequent generation step.
In this step, the model has global, high-level knowledge of the generated data, enabling it to produce a diverse set of subdomains and avoid duplication.
In the second generation step, the model is prompted to generate specific entities for a given subdomain.
Because subdomains are smaller than the full domain, the output size remains manageable, avoiding context window size limitations.
However, since the model lacks awareness of other subdomains' data, we inject factual information, such as globally unique IDs and usernames, directly into the prompt to avoid unintentional overlap between identifiers of different subdomains.

Using this approach, we generated a knowledge base for a university network comprising 50 departments/segments, with a total of 1007 users, 1013 devices, and 571 services.
The numbers and ratios of users, devices, and services per segment were guided by our expert interviews (Section~\ref{sec:foundations:design_principles}) and aligned with those observed in a real university network.
We used GPT-4.1 with a decoding temperature of 1.0 in both generation steps and then parsed the output into an SQL database to verify necessary attribute constraints.

\subsection{Creation of the Request-Intent Pairs}

To make different PCRs more comparable and enable analysis of the impact of different wordings, as might be produced by people from different professional backgrounds, we define classes of PCRs along two dimensions: (1) abstraction level (\texttt{contextless}, \texttt{context-sensitive}, \texttt{interpretive}), and (2) writing style (\texttt{concise}, \texttt{email}, \texttt{cluttered}).
Contextless PCRs mention all necessary attributes (including subject and resource IDs) in the text and can be translated into intents without the knowledge base.
Such PCRs will typically only be written by users with a technical background and a good understanding of the access control system.
Context-sensitive PCRs do not mention subject and resource IDs directly in the text, but refer to them through expressions that can be resolved by consulting the knowledge base (e.g., ``our gitlab server'' could be resolved to resource ID 5).
Interpretive PCRs are those that do not explicitly state the required access control policy change but imply it (e.g., ``gitlab is not working'') and thus represent the highest abstraction level.
Writing styles determine how much additional text is included in the PCR beyond its core statement.
Concise PCRs contain no extra text, email-style PCRs include an additional opening and closing statement, and cluttered PCRs further include a distracting sentence.
A full example of all PCR styles is provided in Appendix~\ref{app:dataset}.

Building on the knowledge base, we then manually constructed pairs of 450 PCRs and their corresponding expected intents.
Five computer network experts, each holding at least an M.Sc. degree in Computer Science and not involved in the system design, were asked to independently develop intent scenarios for the subjects and resources in the knowledge base.
The authors then verified the correctness of all pairs.
All PCR-intent pairs were individually cross-reviewed to ensure that there are no errors in the ground-truth intent, the PCR formulation, or the knowledge base.

NLACBench contains a total of 50 hand-crafted scenarios in three different abstraction levels and three different writing styles (50 $\times$ 3 $\times$ 3 = 450 PCRs in total).
All PCR-intent pairs reference entities within the same five of the total 50 segments of the knowledge base.
This allows us to use the remaining 45 segments to scale the knowledge base in our evaluation, simulating organizations of different sizes.
Positional bias is eliminated during evaluation by shuffling all entities.
\section{Evaluation}
\label{sec:eval}

In this section, we conduct an extensive study to quantify the performance of recent LLMs for intent translation using the NLACBench benchmark dataset.
We further study the effects of system parameters and organization size on accuracy, and the impact of our subgraph construction approach on scalability and costs.
Finally, we analyze failure patterns across models through a case study.
All experimental results, including the full API inputs and outputs, are available alongside our code on GitHub~\cite{artifacts_github} for full reproducibility.

\subsection{Experiment 1 \textendash{} Model Comparison}

In this experiment, we study the intent translation performance of state-of-the-art LLMs.
Here, we measure raw LLM performance (without our subgraph construction technique) to provide an overview of LLMs and select high-performing models for further experiments.
For each Policy Change Request (PCR) in the benchmark dataset, we count the LLM output as correct if it is equal to the expected ground truth intent and compute an accuracy score by averaging over all outputs.
For this and the following experiments, we set the size of the knowledge base to \textbf{five segments} (252 entities) to first analyze the performance of LLMs in small networks.
In our study, we include a broad selection of models, including small- to medium-sized open-source models from different model families that could be run on-premise, as well as OpenAI's proprietary GPT-4.1 models.
Based on preliminary experimental results, we set the number of in-context examples to 1 and the example style to \texttt{context-sensitive} (for details, see Section~\ref{sec:eval:params}).
As in-context examples for a specific PCR, we use other PCRs from the dataset along with their ground truth intents.
Table~\ref{tab:model_overview} presents the experimental results.\footnote{NLACBench has a balanced distribution of PCR abstraction classes. If the distribution for a target organization is known, the total accuracy can be recomputed as a weighted average of the per-abstraction-level accuracies.}

\begin{table}[htb]
  \centering
  \caption{Raw LLM performance on NLACBench for small networks (5 network segments; 252 entities)}
  \label{tab:model_overview}
  \setlength{\tabcolsep}{3pt}
  \resizebox{\linewidth}{!}{%
    \begin{tabular}{lccc|cccc|cc|cc|cc|cc|c|c}
      \toprule
       & \rotatebox{90}{\textbf{GPT-4.1}}      & \rotatebox{90}{GPT-4.1-mini} & \rotatebox{90}{GPT-4.1-nano}
       & \rotatebox{90}{\textbf{Qwen3.5:122b}} & \rotatebox{90}{Qwen3.5:35b}  & \rotatebox{90}{Qwen3.5:9b}   & \rotatebox{90}{Qwen3.5:4b}
       & \rotatebox{90}{GPT-OSS:120b}          & \rotatebox{90}{GPT-OSS:20b}
       & \rotatebox{90}{\textbf{LLaMA3.3:70b}} & \rotatebox{90}{LLaMA3.1:8b}
       & \rotatebox{90}{DS-R1:70b}             & \rotatebox{90}{DS-R1:14b}
       & \rotatebox{90}{gemma3:27b}            & \rotatebox{90}{gemma3:12b}
       & \rotatebox{90}{mistral:7b}
       & \rotatebox{90}{falcon3:10b}                                                                                                      \\
      \midrule
      total\%
       & 96.9                                  & 89.1                         & 45.3
       & 95.8                                  & 90.7                         & 86.0                         & 85.3
       & 86.7                                  & 72.9
       & 83.3                                  & 36.4
       & 64.9                                  & 53.3
       & 59.1                                  & 49.3
       & 32.0
       & 24.2                                                                                                                             \\
      contextless\%
       & 100.0                                 & 99.3                         & 84.0
       & 98.7                                  & 97.3                         & 94.0                         & 96.7
       & 100.0                                 & 96.7
       & 98.7                                  & 72.7
       & 78.0                                  & 86.7
       & 96.0                                  & 90.7
       & 68.7
       & 52.7                                                                                                                             \\
      context-sens.\%
       & 96.0                                  & 86.0                         & 29.3
       & 96.0                                  & 92.7                         & 86.0                         & 82.7
       & 82.0                                  & 62.0
       & 78.7                                  & 27.3
       & 64.7                                  & 44.7
       & 49.3                                  & 28.7
       & 18.7
       & 14.0                                                                                                                             \\
      interpretive\%
       & 94.7                                  & 82.0                         & 22.7
       & 92.7                                  & 82.0                         & 78.0                         & 76.7
       & 78.0                                  & 60.0
       & 72.7                                  & 9.3
       & 52.0                                  & 28.7
       & 32.0                                  & 28.7
       & 8.7
       & 6.0                                                                                                                              \\
      \bottomrule
    \end{tabular}
  }
\end{table}

\noindent
Intent translation capability differs significantly across models, ranging from $24.2$\% to $96.9$\%.
As expected, the level of abstraction of PCR wordings is inversely correlated with model accuracy on these tasks.
Among different writing styles (concise, email, cluttered), we could not measure significant differences, and thus do not show this dimension in the table.
Larger models within the same model family consistently achieve higher accuracy.
To provide a comprehensive analysis of models from different vendors and model families, we select the following models for the subsequent experiments: OpenAI's proprietary model GPT-4.1, Alibaba Cloud's open-source reasoning model Qwen3.5:122b, and Meta's open-source model LLaMA3.3:70b.

\para{Takeaways}
Simple requests that do not involve contextualization or reasoning can reliably be processed by state-of-the-art LLMs, as several models achieve near-perfect accuracy on contextless requests.
However, abstract requests are more challenging and can only be reliably translated by the most capable models.

\subsection{Experiment 2 \textendash{} Ablations} \label{sec:eval:params}

We alter system parameters independently to quantify their impact on intent translation accuracy.

\begin{wrapfigure}{tbh}{0.35\linewidth}
  \centering
  \includegraphics[width=\linewidth]{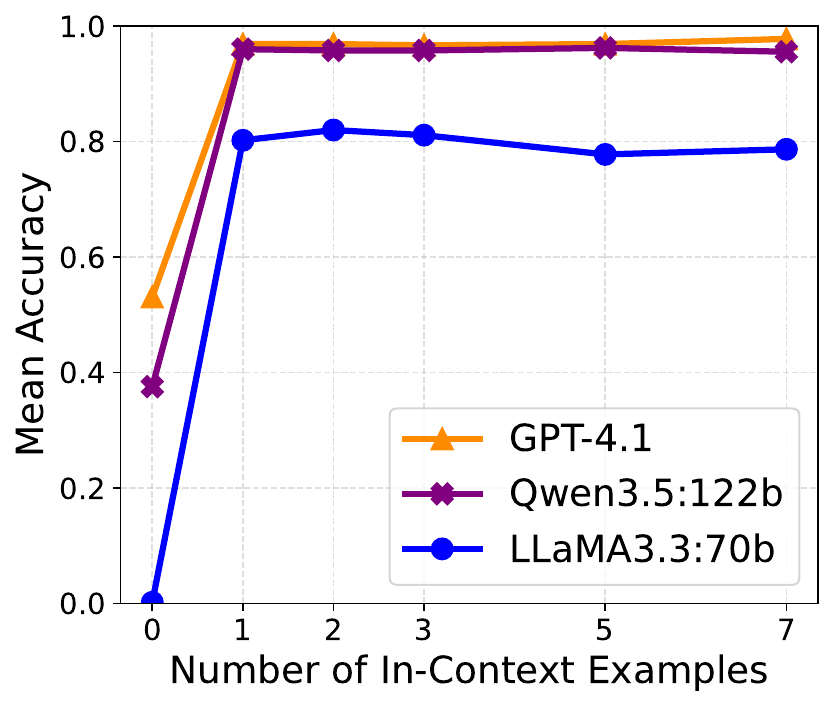}
  \caption{Impact of examples.}
  \label{fig:exp2_2_n_ex}
\end{wrapfigure}

\para{Number of Examples}
The intent translation accuracy using different numbers of in-context examples on NLACBench is shown in Figure~\ref{fig:exp2_2_n_ex}.
We find that a single in-context example is sufficient to significantly improve task performance.
Additional examples do not lead to further accuracy gains, suggesting that the benefit of examples lies more in demonstrating the general format (e.g., input and output structure) than in their domain-specific content.

\para{Style of Examples}
We experiment with varying the abstraction level of the provided examples (\texttt{contextless}, \texttt{context-sensitive}, \texttt{interpretive}).
For each example type, we provide a Chain-of-Thought~(CoT)~\cite{wei2022chain} to the solution.
We could not measure a significant effect across these variations, neither in terms of total aggregated accuracy nor for requests of specific abstraction levels.
This shows that in-context learning from reasoning chain demonstrations is not effective for intent translation.

\headlesspara
With these insights about system parameter impacts, we set the number of examples to 1 and example strategy to \texttt{context-sensitive} for the subsequent experiments.

\para{Takeaways}
In-context examples for intent translation are an effective way to improve LLM alignment to human expectations.
However, they appear to offer no benefit for domain-specific problem understanding and reasoning.

\subsection{Experiment 3 \textendash{} Scaling to Larger Networks} \label{sec:eval:exp3}

In the previous experiments, we analyzed the performance of LLMs for intent translation in the context of small networks, using only five of the 50 knowledge base segments, accounting for 252 entities.
In this experiment, we study the scalability of LLM-based intent translation to larger networks.
We further evaluate our semantic subgraph construction approach with respect to its effectiveness in improving scalability and reducing operating costs of intent translation.

\begin{figure}[h]
  \centering
  \begin{subfigure}[t]{0.4\linewidth}
    \centering
    \includegraphics[width=\linewidth]{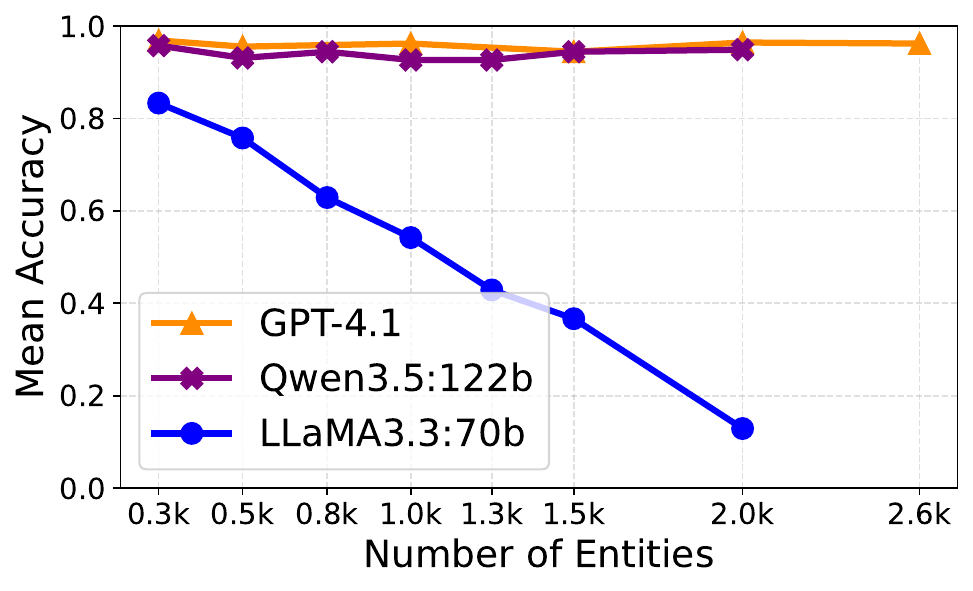}
    \vspace{-0.6cm}
    \caption{Standalone LLM.}
    \label{fig:exp3:baseline}
  \end{subfigure}
  \hspace{0.03\linewidth}
  \begin{subfigure}[t]{0.4\linewidth}
    \centering
    \includegraphics[width=\linewidth]{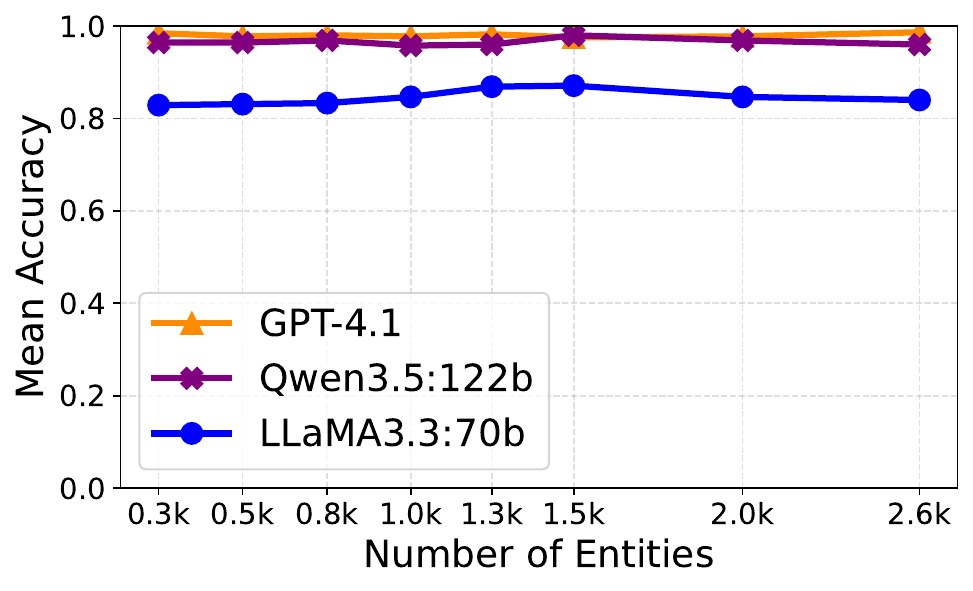}
    \vspace{-0.6cm}
    \caption{Subgraph construction context filter.}
    \label{fig:exp3:filtering}
  \end{subfigure}
  \vspace{-0.2cm}
  \caption{Scalability on NLACBench from 5 to 50 network segments (252 to 2591 entities).}
  \label{fig:exp3}
\end{figure}

\noindent
To test the intent translation accuracy in larger organizations, we gradually increase the size of the knowledge base from five to 50 segments (up to 2591 entities), as shown in Figure~\ref{fig:exp3}. For LLaMA3 and Qwen3.5, we stopped at 40 segments due to hardware and context window limitations.
For the baseline configurations without subgraph construction (Figure~\ref{fig:exp3:baseline}), GPT-4.1 and Qwen3.5:122b do not experience significant accuracy drops for larger organizations, while LLaMA3.3:70b shows poor scalability with a linear decline in accuracy.
With subgraph construction turned on (Figure~\ref{fig:exp3:filtering}), we find that the performance of LLaMA3.3:70b remains stable, yielding an accuracy improvement from below 20\% to over 80\% for large organizations (40+ segments).
For GPT-4.1 and Qwen3.5, subgraph construction has a minor positive impact on accuracy (1-2 percentage points).

\begin{figure}[h]
  \centering
  \begin{subfigure}[t]{0.4\linewidth}
    \centering
    \includegraphics[width=\linewidth]{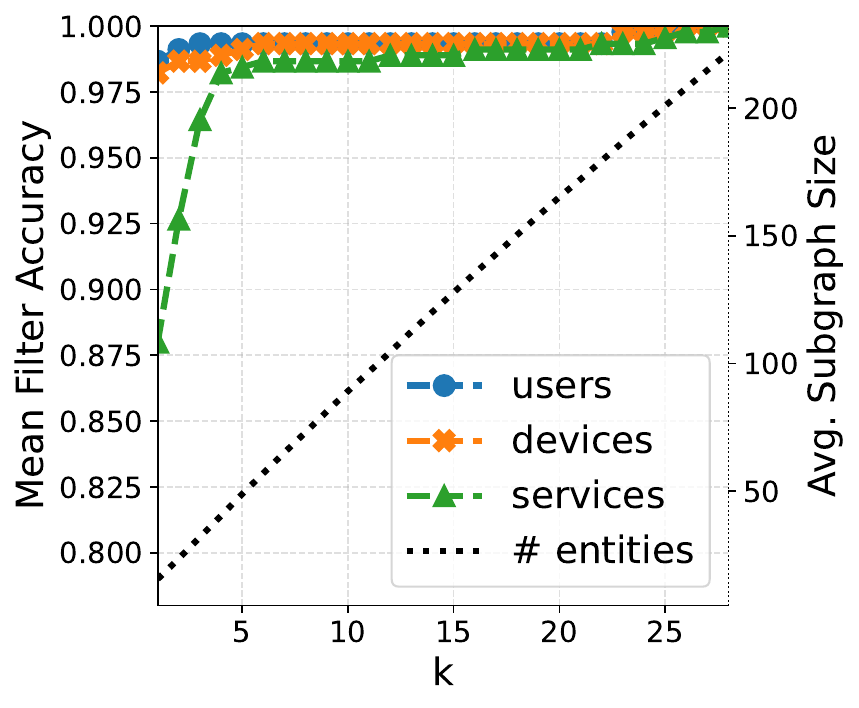}
    \vspace{-0.6cm}
    \caption{Retrieval accuracy}
    \label{fig:exp4_filter}
  \end{subfigure}
  \hspace{0.03\linewidth}
  \begin{subfigure}[t]{0.4\linewidth}
    \centering
    \includegraphics[width=\linewidth]{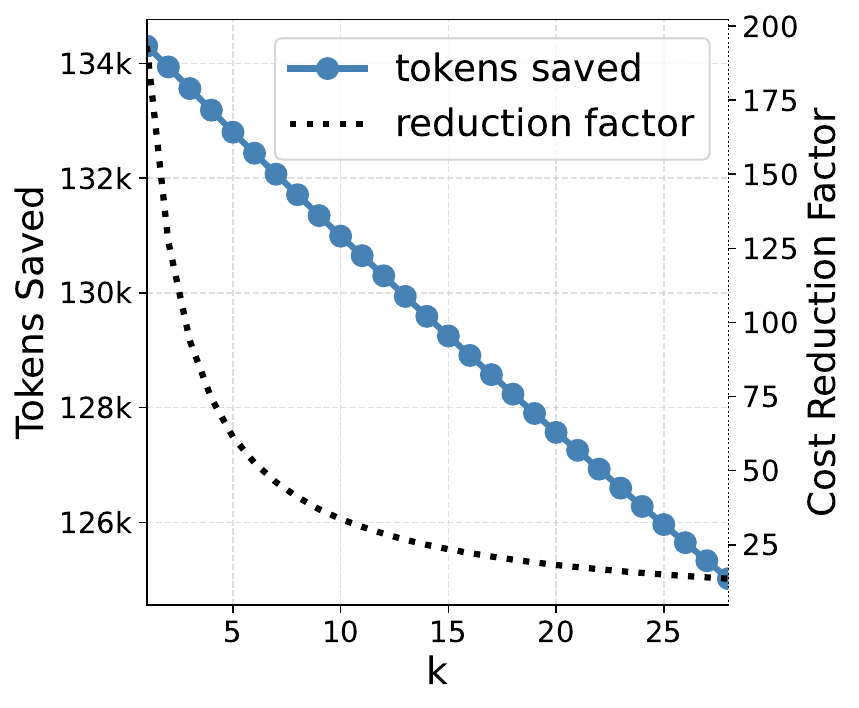}
    \vspace{-0.6cm}
    \caption{Input token savings.}
    \label{fig:exp4_tokens}
  \end{subfigure}
  \vspace{-0.2cm}
  \caption{Optimizing the number of retrieved entities $k$ for semantic subgraph construction.}
  \label{fig:exp4}
\end{figure}

To optimize the number of nearest neighbors ($k$) retrieved for the subgraph, we analyzed the subgraph construction accuracy and the cost savings for different $k$ values on the largest benchmark configuration (50 segments, 2591 entities), as shown in Figure~\ref{fig:exp4}.
We rate a sample as correct if the subject and the resource of the expected intent are contained in the built subgraph.
We find that our subgraph construction approach is highly efficient:
For a given PCR, in over 97.5\% of cases, it requires looking up only the 4 nearest neighbors in the embedding space to build a network subgraph that contains all relevant entities.
Using $k=4$, our approach saves 133k input tokens, representing an input token reduction by a factor of 75x.
Using $k=28$, we achieve 100\% filter accuracy (for all PCR in the benchmark dataset, the subgraph always contains all relevant entities for this PCR), while still saving 125.5k input tokens, reducing input token costs by a factor of over 24x.
For the scalability experiments shown in Figure~\ref{fig:exp3}, we selected the minimum number of entities that achieves 100\% filter accuracy per entity type: $k_{\text{users}}=26$, $k_{\text{devices}}=26$, and $k_{\text{services}}=28$.

\para{Implications}
Besides significantly improving intent translation performance for some LLMs (from below 20\% to over 80\% for LLaMA3.3:70b), the reduction of input size also has operational implications:
Per-request computation time is reduced, leading to reduced GPU resource consumption, for instance from 7:00 min to 12 sec for LLaMA3.3:70b; and from 7:30 min to 52 sec for Qwen3.5:122b.
This also affects commercial API inference costs (usually billed per token), as, for example, API costs for GPT-4.1 are reduced to 2.9\% of the original costs.

Furthermore, VRAM requirements are reduced:
For LLaMA3.3:70b, the VRAM footprint drops from 105 GB to 45 GB; and for Qwen3.5:122b from 133GB to 89 GB.
This may enable deploying intent translation systems on locally hosted infrastructure and mitigate privacy concerns associated with sharing confidential data with third-party API providers.

Lastly, reducing the input token count to a constant number overcomes LLM fixed context window limitations (e.g., LLaMA3.3:70b's 132k context window can only fit small to medium-sized knowledge bases). This broadens the available options for LLMs to choose from, which is particularly relevant to stay independent of individual LLM providers and maintain digital sovereignty of security-critical systems.

\para{Takeaways}
Network graph filtering can significantly reduce costs to a constant budget without impairing translation accuracy, and for some models substantially improves it.

\subsection{Experiment 4 \textendash{} Case Study: Error Analysis} \label{sec:eval:casestudy}

To characterize residual failure modes of the two top-performing models, we identify benchmark cases on which at least one model exhibits systematic errors and re-run inference with $n=100$ different random seeds and temperature set to 1.
We analyze two formulations of the same use case: a \texttt{context-sensitive} variant that names entities explicitly and an \texttt{interpretive} variant that conveys the request indirectly.
In both cases, both models consistently resolve the correct subject and action; errors concern only the target service and source device constraint, as summarized in Table~\ref{tab:casestudy}.

{\Small
\begin{table}[htb]
  \centering
  \caption{Accuracy and dominant error type over $n{=}100$ random seeds for one use case in NLACBench.}
  \label{tab:casestudy}
  \setlength{\tabcolsep}{5pt}
  \begin{tabular}{llr|lr}
    \toprule
    \textbf{PCR}                                                       & \textbf{Model} & \textbf{Acc.} & \textbf{Primary error} & \textbf{Freq.} \\
    \midrule
    \emph{``Could you please give Vincent Parks ssh access from his}    & GPT-4.1        & 99\%          & –          & 1\%            \\
    \emph{PC to the server for code deployment?''} & Qwen3.5:122b   & 20\%          & wrong service          & 73\%           \\
    \midrule
    \emph{``Our Lab Technician has written some programs on his PC}    & GPT-4.1        & 38\%          & constraint omitted        & 60\%           \\
    \emph{that he wants to deploy to the deployment server via SSH.''} & Qwen3.5:122b   & 78\%          & wrong service          & 20\%           \\
    \bottomrule
  \end{tabular}
\end{table}
}

\para{Context-Sensitive PCR}
GPT-4.1 resolves all named entities correctly in $99\%$ of runs.
Qwen3.5:122b achieves only $20\%$ accuracy: in $73\%$ of runs it maps the destination to a different SSH-accessible service, indicating that semantically similar service descriptions mislead its entity disambiguation.

\para{Interpretive PCR}
Performance is inverted.
Qwen3.5:122b reaches $78\%$ accuracy, correctly inferring the user's PC as implied execution context constraint.
GPT-4.1 accuracy drops to $38\%$: in $60\%$ of runs it resolves the user and service correctly but omits the execution context constraint.

\para{Implications}
This analysis indicates that GPT-4.1 excels at entity disambiguation but tends to miss implicit scope constraints, while Qwen3.5:122b handles indirect reasoning better but is more susceptible to ambiguities in context resolution.

\para{Takeaways}
Our case study indicates that top-performing models exhibit partially complementary failure modes, motivating further investigation into multi-LLM approaches to leverage their combined strengths.

\section{Discussion}

Separating LLM-based intent translation from rule-based configuration updating in the Natural Language Access Control (NLAC) architecture keeps the scope of intent translation manageable for the LLM while ensuring that outputs are compact and efficient to verify.
Our experiments on NLACBench show that intent translation performance differs substantially across LLMs and that intent translation poses a significant challenge to most tested LLMs.
A few top-performing LLMs achieve reasonably high accuracy (over 95\%), suggesting potential for substantial automation of access control management.
Yet, scaling our experiments to larger networks revealed degraded performance as well as increased hardware requirements and costs.
Filtering the network graph based on semantic similarity appears to be an effective approach to address these limitations.
This improvement could enable the use of on-premises hosted LLMs for intent translation on less expensive hardware, thereby avoiding sharing sensitive data with third-party API providers.
Partially complementary failure patterns in high-performing reasoning (Qwen3.5:122b) and non-reasoning (GPT-4.1) LLMs from different providers motivate future work on multi-LLM cross-checking approaches to combine the strengths of different models.

\para{Limitations}
First, NLACBench is designed to resemble university networks.
Although results on NLACBench may generalize to other settings, such as enterprise networks, there could be differences in the characteristics of such networks that are not modeled in NLACBench.
Second, adversarial examples or unsolvable requests are considered out of the scope of NLACBench, as we aim to evaluate performance rather than robustness.
However, undesired requests or LLM outputs can be caught by human review, which could be complemented with machine learning-based detection mechanisms.
Third, our evaluation assumes the availability of a high-quality knowledge base.
If such a knowledge base is not available, it could be automatically synthesized from documentation, websites, and network scans.
Finally, the impact of NLAC on human operator efficiency in real-world deployments remains subject to further research.

\para{Ethical Considerations}
We provide a detailed stakeholder-based ethics analysis in Appendix~\ref{app:ethics}. In summary, our work raises no severe ethical concerns.

\section{Security Implications}\label{sec:security}

\para{Threat Model}
We consider two realistic adversaries, $A_1$ and $A_2$.
$A_1$ may send arbitrary requests to the intent translation module, including unreasonable ones or those containing prompt injection attacks.
$A_2$ may modify the knowledge base contents. This would be possible if the knowledge base were synthesized from insufficiently protected information sources, such as documentation and websites.

\headlesspara
Attacker $A_1$ may cause the intent translation module to output arbitrary structured intents $(\sigma, \delta)$, which are subsequently reviewed.
Outputs that do not conform to the formal specification of an intent do not pose a risk, as they will result in parsing errors.
We argue that the risk posed by attacker $A_1$ is no different from the risk in classical policy management workflows, where users can send unreasonable or malicious requests to the help desk.
In both cases, humans need to decide whether to implement the requested changes.
Whether the Natural Language Access Control (NLAC) architecture influences the performance of human reviewers remains subject to future research.

\headlesspara
As with $A_1$, attacker $A_2$ may in the worst case also cause the intent translation module to output arbitrary intents by injecting malicious data into the LLM context.
Additionally, if the reviewer also relies on the knowledge base as context for the review, the attacker may be able to deceive them.
Despite this risk, we argue that the situation is not fundamentally different from classical policy management workflows:
A similar risk exists if help desk staff rely on compromised information, such as incorrect user IDs or false identities, to formulate access policies.

\section{Conclusions}

In this work, we proposed Natural Language Access Control (NLAC), an architecture and policy model for intent-based access control that provides a natural language interface for end users.
The architecture leverages LLMs to interpret help-desk-like requests, while handling policy updates with rule-based algorithms and keeping deployment decisions under human control.
To evaluate the capabilities of LLM-based intent translation approaches, we created and published a benchmark dataset, named NLACBench.
Experimental results on NLACBench indicate that many popular LLMs struggle with intent translation, while a small number of recent high-capability LLMs show strong potential, achieving high benchmark performance ($>$95\% accuracy).
However, we also identify scalability limitations of top-performing models, which we address by filtering LLM inputs based on semantic similarity.
Using this approach, LLM-based intent translation can scale to large networks using locally hosted open-source models.
The results of our case study indicate that failure patterns in top-performing models differ and are partially complementary, suggesting the potential for multi-LLM approaches to achieve even higher accuracy.
Overall, natural language access control could provide substantial benefits for automating network access control management.

\bibliographystyle{ACM-Reference-Format}
\bibliography{references,references_web}

\appendix

\cleardoublepage

\section{Ethical Considerations}\label{app:ethics}

We provide a stakeholder-based ethics analysis of the potential impacts of our work.

\para{Network Operations Teams}
Our work can reduce the workload of network operations teams, allowing them to focus on higher-level tasks.
A possible downside is reduced demand for positions focused specifically on policy management.

\para{Parties Involved with Our Requirements Analysis}
The set of real-world help desk inquiries used for requirements engineering was accessible to co-authors through their professional responsibilities at a compute center.
No institution-specific information was disclosed, included in the derived requirements, mentioned in the paper, incorporated into the benchmark dataset, or disclosed in any other form.
Likewise, interviews with network operators were used solely for requirements elicitation, and no sensitive information was disclosed or included in our work.

\para{Environmental Impact}
We designed our experiments to balance statistical validity with environmental impact.
Initial experiments narrowed the set of LLMs used in later stages, avoiding unnecessary runs of large models.
In production, the environmental impact is expected to remain low because organizations typically generate only a few requests per day, each requiring a single LLM call.
Our semantic subgraph construction method further reduces LLM input size, lowering computation time and energy consumption.

\para{LLM and Computer Hardware Providers}
We recognize that adoption of our approach may increase dependence on LLM API and hardware providers.
To mitigate this dependency, we studied the use of open-source LLMs that can be deployed and operated on-premise without third-party providers.
\section{Artifacts}

All code, data, and results used in this work will be made publicly available under an MIT License to serve as a basis for future research. The list of artifacts includes:

\begin{itemize}[]
    \item NLACBench: The benchmark dataset proposed in this work and used for our evaluation,
    \item intent translation source code: Our source code for the intent translation system evaluated on NLACBench,
    \item experimental results: All experimental results of this work, including all inputs and outputs to APIs for full reproducibility.
\end{itemize}
\section{Example: Processing of Change Requests}\label{app:processing_of_pcrs}

In the following, we provide a walkthrough of an example illustrating how a natural-language user request is converted into a structured representation and integrated into the access control configuration according to the NLAC architecture (see Section~\ref{sec:architecture}).

\subsection{Setting}

For this simple example, let the knowledge base contain the following information:
\begin{itemize}[noitemsep]
    \item A list of all users, among which one user has the ID 101.
    \item A list of services, among which one has the title ``Robot Arm Controller'', the ID 107, and runs an encrypted REST endpoint on port 8000.
\end{itemize}

\noindent
Let the schema for Resource Access Requests (RARs) be:

$$
    r = (a_{\text{sub}}, a_{\text{res}}, a_{\text{port}}, a_{\text{prot}})
$$

\noindent
with

\begin{itemize}[noitemsep]
    \item $a_{\text{sub}}, a_{\text{res}} \in \mathbb{N}_{0}$ denoting the subject and the resource IDs, respectively,
    \item $a_{\text{port}} \in \{n \in \mathbb{N}_{0} \; | \;  n \leq 65535\}$ denoting the destination port number, and
    \item $a_{\text{prot}} \in \{n \in \mathbb{N}_{0} \; | \;  n \leq 255\}$ denoting the IP protocol number as defined by the Internet Assigned Numbers Authority (IANA).
\end{itemize}

\noindent
Let the current state of the Intent Policy Configuration (IPC) (see Section~\ref{sec:architecture:ipc}) be defined as follows:

\begin{itemize}[noitemsep]
    \item There exist two administrative groups, $g_1$ and $g_2$. The user with ID 101 belongs to group $g_1$, while the service with ID 107 belongs to group $g_2$, meaning they are managed by different policy owners.
    \item $\Pi^\mathrm{in}_1 = \varnothing$ and $\Pi^\mathrm{out}_1 = \{((a_{\text{sub}} \to a_{\text{sub}} = 101), (a_{\text{res}} \to \text{true}), (a_{\text{port}} \to \text{true}), (a_{\text{prot}} \to \text{true}))\}$ represent the current configuration of the ingress and egress policy sets for group $g_1$, meaning that the user with ID 101 is allowed to make outgoing requests from their group, while no incoming requests are allowed for this group. Note that for improved readability we use $(a \to \text{expr})$ as a shorthand for $(a \to \text{true if expr else false} )$ 
    \item $\Pi^\mathrm{in}_2 = \{((a_{\text{sub}} \to a_{\text{sub}} = 42), (a_{\text{res}} \to a_{\text{res}} = 107), (a_{\text{port}} \to a_{\text{port}} = 8000), (a_{\text{prot}} \to a_{\text{prot}} = 6))\}$ and $\Pi^\mathrm{out}_2 = \varnothing$ represent the current configuration of the ingress and egress policy sets for group $g_2$, meaning that incoming requests from a user with ID 42 via protocol 6 (TCP) on port 8000 are allowed, while no outgoing requests are allowed for this group.
\end{itemize}

\noindent
Let us further assume that the user with ID 101 submits the following Policy Change Request (PCR) to the NLAC system:\\``I need to do experiments with the robot arm.''

\subsection{Intent Translation}

In the intent translation stage, the intent translation system will receive the PCR and the knowledge base as inputs.
It will interpret the PCR within the context of the knowledge base, find the targeted robot arm controller REST endpoint in the knowledge base, identify important attributes such as port and protocol, and construct the following intent:

$$
    (\sigma, \delta) =
    \left(
    \begin{array}{l}
            (a_{\text{sub}} \to a_{\text{sub}} = 101),    \\
            (a_{\text{res}} \to a_{\text{res}} = 107),    \\
            (a_{\text{port}} \to a_{\text{port}} = 8000), \\
            (a_{\text{prot}} \to a_{\text{prot}} = 6)
        \end{array},
    \, 1
    \right),
$$

\noindent
which will be passed to the intent resolution stage.

\subsection{Intent Resolution}

The intent is now integrated into the current state of the IPC.
We first determine the relevant policy sets, which are $\Pi_{1}^{out}$ and $\Pi_{2}^{in}$ in this case.
Both of these policy sets could potentially be affected by the intent.
For each of these sets, we now compute the proposed updates according to Algorithm~\ref{algo:update_sets}:

\begin{itemize}[noitemsep]
    \item \texttt{UpdateSelectorSets(}$(\sigma, \delta), \Pi_{1}^{out}\texttt{)}$: Since $\delta$ is equal to 1, we check whether there is a selector in $\Pi_{1}^{out}$ that covers $\sigma$. We find that such a selector exists, since the selector in the selector set matches all RAR attributes that $\sigma$ matches. Consequently, $\sigma$ is already implied by the selector set $\Pi_{1}^{out}$ and the selector set does not have to be modified.
    \item \texttt{UpdateSelectorSets(}$(\sigma, \delta), \Pi_{2}^{in}\texttt{)}$: Since $\delta$ is equal to 1, we check whether there is a selector in $\Pi_{2}^{in}$ that covers $\sigma$. We find that there is no such selector, since the only selector present in $\Pi_{2}^{in}$ and $\sigma$ match disjoint subject IDs. Then, we add $\sigma$ to $\Pi_{2}^{in}$, leading to $\Pi_{2}^{in} = \{((a_{\text{sub}} \to a_{\text{sub}} = 42), (a_{\text{res}} \to a_{\text{res}} = 107), (a_{\text{port}} \to a_{\text{port}} = 8000), (a_{\text{prot}} \to a_{\text{prot}} = 6)), ((a_{\text{sub}} \to a_{\text{sub}} = 101), (a_{\text{res}} \to a_{\text{res}} = 107), (a_{\text{port}} \to a_{\text{port}} = 8000), (a_{\text{prot}} \to a_{\text{prot}} = 6))\}$.\\
    Optionally, this expression could be simplified to $\Pi_{2}^{in} = \{((a_{\text{sub}} \to a_{\text{sub}} \in \{42, 101\}), (a_{\text{res}} \to a_{\text{res}} = 107), (a_{\text{port}} \to a_{\text{port}} = 8000), (a_{\text{prot}} \to a_{\text{prot}} = 6))\}$.
\end{itemize}

\subsection{Change Authorization}

Now that the necessary changes to the selector sets have been determined, the policy owners of the administrative groups will be prompted to authorize the changes.
For group $g_1$, the policy owner will not have to authorize any changes, since their policy sets do not need to change.
For group $g_2$, the policy owner will have to decide whether to authorize or decline the proposed changes to their ingress policy set, i.e., they will have to decide whether they want to allow the new access policy to their REST endpoint.
To display the diff in a more user-friendly way, we can resolve IDs to descriptive labels:\\

{\centering
\usetikzlibrary{positioning, shapes.geometric, arrows.meta, fit}
\begin{tikzpicture}[
        smallbox/.style={font=\small,rectangle, fill=green!10, draw, rounded corners, minimum width=3cm, align=center},
    ]
    tikzset{
            smallbox/.style={
                    draw, rounded corners, fill=gray!10,
                    align=left,
                    text width=\linewidth,
                    inner sep=3mm,
                }}

    \node[smallbox] (prompt) {
        \begin{minipage}{0.5\linewidth}
            \setlength{\parskip}{0.5em}
            \textbf{Access Request}
            <user name> requests access to the `Robot Arm Controller` endpoint via TCP on port 8000.
            \begin{tabular*}{\linewidth}{@{\extracolsep{\fill}}c c}
                \framebox[3cm]{Approve} & \framebox[3cm]{Decline\phantom{p}}
            \end{tabular*}

        \end{minipage}

    };
\end{tikzpicture}
}

\subsection{Decision Making on the Data Plane}

Let us now assume that the policy owner of group $g_2$ has authorized the changes.
Let us further assume that the user with ID $101$ now tries to access the robot arm controller REST endpoint.
This request will result in the following RAR:

$$
    r = (101, 107, 8000, 6)
$$
When this RAR is issued, the policy enforcement point will consult the policy engine for decision making (see Figure~\ref{fig:architecture}).
The policy engine, in turn, will use the current state of the Intent Policy Configuration (IPC) (Section~\ref{sec:architecture:ipc}) to decide whether the RAR will be allowed.
An RAR is granted if there is at least one selector in the egress policy set of the subject that matches the RAR, and likewise, at least one selector in the ingress policy set of the resource that matches the RAR.

To check this condition, we start with $\Pi_{1}^{out}$ and find that $((a_{\text{sub}} \to a_{\text{sub}} = 101), (a_{\text{res}} \to \text{true}), (a_{\text{port}} \to \text{true}), (a_{\text{prot}} \to \text{true})) \; \triangleright \; r$ because each predicate in the selector returns true for the corresponding attribute of $r$. Analogously, when checking $\Pi_{2}^{in}$, we find that $((a_{\text{sub}} \to a_{\text{sub}} = 42), (a_{\text{res}} \to a_{\text{res}} = 107), (a_{\text{port}} \to a_{\text{port}} = 8000), (a_{\text{prot}} \to a_{\text{prot}} = 6)) \; \triangleright \; r$.
Thus, we conclude that $r$ is allowed under the current state of the IPC, and the RAR will be allowed by the policy engine, which is then enforced by the policy enforcement point.
\section{Additional Formal Definitions}\label{app:exclude_def}

The $\mathrm{exclude}(\sigma', \sigma)$ subroutine in Algorithm~\ref{algo:update_sets} computes the set-theoretic difference of two selectors' coverages, returning the result as a concrete set of selectors.

We model each selector $\sigma$ as a tuple $(S_1, \dots, S_n)$ of attribute value sets over the RAR schema $A_1 \times \dots \times A_n$, where $\sigma \triangleright r \Leftrightarrow \forall\, i \colon a_i \in S_i$.
Given selectors $\sigma' = (S'_1, \dots, S'_n)$ and $\sigma = (S_1, \dots, S_n)$, the subroutine returns a set of selectors $\Sigma$ satisfying:
\[
  \forall\, r \in A_1 \times \dots \times A_n \colon
  \quad (\exists\, s \in \Sigma \colon s \triangleright r) \;\Longleftrightarrow\; (\sigma' \triangleright r) \land \lnot(\sigma \triangleright r)
\]
One concrete construction decomposes this into at most $n$ disjoint selectors:
\[
  \mathrm{exclude}(\sigma', \sigma) = \bigl\{\, \sigma_i \;\big|\; i \in \{1,\dots,n\},\; S'_i \setminus S_i \neq \emptyset \,\bigr\}
\]
where the $j$-th component of each $\sigma_i$ is:
\[
  (\sigma_i)_j = \begin{cases}
    S'_j \cap S_j & \text{if } j < i \quad \text{($r$ still inside $\sigma$)} \\
    S'_j \setminus S_j & \text{if } j = i \quad \text{($r$ escapes $\sigma$ here)} \\
    S'_j & \text{if } j > i \quad \text{(free within $\sigma'$)}
  \end{cases}
\]
Each RAR belongs to exactly one $\sigma_i$ (the slice at which it first escapes $\sigma$), so the selectors are pairwise disjoint and their union covers exactly $\sigma' \setminus \sigma$.
In the NLAC context, attribute sets represent possible user IDs, IP address ranges, port ranges, or protocol identifiers, for which the required set operations (complement and intersection) are straightforward to compute.

\section{Abstraction Levels and Writing Styles in NLACBench}\label{app:dataset}

\begin{figure}[H]
    \centering
    \usetikzlibrary{positioning, shapes.geometric, arrows.meta, fit}
    \begin{tikzpicture}[
            smallbox/.style={font=\small,rectangle, fill=green!10, draw, rounded corners, minimum width=3cm, align=center},
        ]
        tikzset{
                smallbox/.style={
                        draw, rounded corners, fill=gray!10,
                        align=left,
                        text width=0.95\linewidth,
                        inner sep=3mm,
                    }}

        \node[smallbox] (concise) {
            \begin{minipage}{0.95\linewidth}
                \textbf{concise:} Could you please give <name> access to the Alumni Mentorship Platform.
            \end{minipage}
        };

        \node[smallbox, below=0.1cm of concise] (email) {
            \begin{minipage}{0.95\linewidth}
                \textbf{email:} Dear network support team,\\Could you please give <name> access to the Alumni Mentorship Platform.\\\
                {Thank you very much!} Best regards
            \end{minipage}
        };
        \node[smallbox, below=0.1cm of email] (cluttered) {
            \begin{minipage}{0.95\linewidth}
                \textbf{cluttered:} Dear network support team,\\
                I hope you enjoyed your holidays! In the meanwhile, we continued working on our current project. Could you please give <name> access to the Alumni Mentorship Platform.\\
                Thank you very much! Best regards
            \end{minipage}
        };
    \end{tikzpicture}
    \caption{Examples of policy change requests in different writing styles in NLACBench.}
\end{figure}

\begin{figure}[H]
    \centering
    \usetikzlibrary{positioning, shapes.geometric, arrows.meta, fit}
    \begin{tikzpicture}[
            smallbox/.style={font=\small,rectangle, fill=green!10, draw, rounded corners, minimum width=3cm, align=center},
        ]
        tikzset{
                smallbox/.style={
                        draw, rounded corners, fill=gray!10,
                        align=left,
                        text width=0.95\linewidth,
                        inner sep=3mm,
                    }}

        \node[smallbox] (contextless) {
            \begin{minipage}{0.95\linewidth}
                \textbf{contextless:} Please allow https traffic from user with id 405 to service with id 405.
            \end{minipage}
        };

        \node[smallbox, below=0.1cm of contextless] (contextsentitive) {
            \begin{minipage}{0.95\linewidth}
                \textbf{context-sensitive:} Could you please give <name> access to the Alumni Mentorship Platform.
            \end{minipage}
        };
        \node[smallbox, below=0.1cm of contextsentitive] (interpretive) {
            \begin{minipage}{0.95\linewidth}
                \textbf{interpretive:} <name> is helping us to update the information on the mentorship platform. Could you please give them access.
            \end{minipage}
        };
    \end{tikzpicture}
    \caption{Examples of policy change requests at different abstraction levels in NLACBench.}
\end{figure}
\section{Prompt Template}\label{app:prompts}

\begin{figure}[H]
    \centering
    \usetikzlibrary{positioning, shapes.geometric, arrows.meta, fit}
    \begin{tikzpicture}[
            smallbox/.style={font=\small,rectangle, fill=green!10, draw, rounded corners, minimum width=3cm, align=center},
        ]
        tikzset{
                smallbox/.style={
                        draw, rounded corners, fill=gray!10,
                        align=left,
                        text width=\linewidth,
                        inner sep=3mm,
                    }}

        \node[smallbox] (prompt) {
            \begin{minipage}{0.95\linewidth}
                \setlength{\parskip}{0.5em}
                \textbf{[System]:}
                You are an IT help desk assistant at a university. You assist with translating end-user requests into well-defined access policies. You respond in valid JSON only. You will receive an end-user request that asks for allowing or denying network access. User requests are sometimes vague, and you will need to use the provided context to infer the identifiers for the access policy that addresses their request.

                $<$JSON Format Specification$>$

                \textbf{[User]:} $<$Example Request 1$>$\\
                \textbf{[Assistant]:} $<$Example Solution 1$>$

                (... further in-context examples)

                \textbf{[User]:} Please consider the following context for translating the request:\\
                $<$Context Information$>$ \\
                $<$Policy Change Request$>$

                Please extract the structured policy from the user request.
            \end{minipage}

        };
    \end{tikzpicture}
    \caption{Prompt template used in our experiments.}
\end{figure}

\end{document}